\documentstyle[12pt,aaspp4]{article}

\lefthead{Knezek and Bregman}
\righthead{Identification of Quasars}

\begin{document}

\title{The Identification of Quasars Behind Elliptical Galaxies and Clusters of
Galaxies}

\author{Patricia M. Knezek \altaffilmark{1}}
\affil{Observatories of the Carnegie Institution of Washington, Las Campanas Observatory, Casilla 601, La Serena, Chile \\ Electronic mail: pmk@pha.jhu.edu}
\and 
\author{Joel N. Bregman} 
\affil{Department of Astronomy, University of Michigan, Ann Arbor, MI 48109-1090\\ Electronic mail: jbregman@astro.lsa.umich.edu}
\authoraddr{Las Campanas Observatory, Casilla 601, La Serena, Chile}
\altaffiltext{1}{Now at the Department of Physics and Astronomy, The Johns Hopkins University, Baltimore, MD 21218-2695}

\begin{abstract}
The detection of resonance absorption lines against known objects
such as individual galaxies and clusters of galaxies 
is a powerful approach for studying the gas content of these
systems.  We describe an efficient method of identifying
background quasars suitable for absorption line studies.  In this
finding technique, we identify serendipitous X-ray sources, about
1/8 of which are suitably bright quasars at moderate redshift. 
We identify 16 new quasars and galaxies with active galactic nuclei (AGNs),
and confirm 5 known quasars and AGNs,  
superimposed behind elliptical galaxies and clusters of galaxies.  We also
present 3 QSO/AGN candidates with uncertain redshift identifications.
\end{abstract}


\section{Introduction}
Our present understanding of the gaseous content of elliptical
galaxies and of clusters of galaxies results largely from studies
of the emission of photons from ionized plasmas and excited
atoms. Studies of emission require that gas of particular
temperature have an adequate emission measure, so it is
insensitive to even large amounts of material for which the
species are in their ground state.  However, weakly excited gas
can produce absorption lines against background continuum
sources, so this gas can be detected, and with considerable
sensitivity.  In order to search for the absorbing gas, we have
developed a strategy to identify background continuum sources, mostly
quasars, behind elliptical galaxies and clusters of galaxies. 
The identification of this absorbing material not only improves
our census of the interstellar content of these systems, it also
tests several astrophysical models for its origin.

     A variety of astrophysical events can occur in ellipticals
and in clusters of galaxies that are unlikely to have a
detectable emission signature but may lead to detectable
absorption lines.  For example, cooling and cooled gas is
anticipated from the radiative losses (the X-rays) of the hot
interstellar gas in these systems (e.\ g., from cooling flows, 
Fabian, Nulsen, and Canizares 1991; see also reviews by Fabbiano
1989 and Sarazin 1990). 
Two other sources for absorbing gas in elliptical galaxies are
mass shed from evolving stars and accretion of gas onto the
galaxy (Mathews and Baker 1971, Bregman 1978, White and Chevalier 1983).  
Important sources of absorbing gas in clusters of
galaxies are from gas stripped out of galaxies (Gaetz, Salpeter, and 
Shaviv 1987) and infall of
material into the cluster (Metzler 1995).  The properties of the absorbing gas
are often different for the various processes, so the detection
of absorption lines can potentially identify the important processes in these
systems.

     Although many quasars are known, galaxies and clusters
subtend a small solid angle on the sky, so there are relatively
few coincidences with quasars.  (Given the redshift range that we are
searching for backgound objects, {z $\sim$ 0.1--1, we choose to use the 
terminology ``quasar'' in this paper, rather than active galactic nuclei 
(AGN) to designate emission line candidates.)
Furthermore, most searches for quasars and
active galactic nuclei avoided the region near bright galaxies. 
Therefore, there is a need to identify a reasonable number of new
quasars near ellipticals and clusters of galaxies.  These quasars
should be bright enough to permit absorption line observations to
be made with present telescopes, such as the Hubble Space
Telescope (HST), in a reasonable amount of time (most of the quasars
identified were scheduled for HST observations).  Here we describe
an effective technique for finding background quasars that makes
efficient use of telescope time.

\section{The Quasar Search Technique}
     It is valuable to understand the type of quasar needed for
absorption line experiments in order to optimize a search
technique.  The most effective searches make use of resonance
line absorption, and for the optical region, we are limited to
\ion{Na}{1} and \ion{Ca}{2}.  Not only are these uncommon species, but under
many conditions, these will be poorly populated ionization
states.  Many of the most common elements, such as hydrogen,
carbon, nitrogen, oxygen, iron, and magnesium, have resonance
absorption lines in the ultraviolet region accessible by the
HST.  When one considers the abundance of the
elements, the ionization state of the species, the spectral shape
of quasars and the wavelength sensitivity of HST, the low
ionization state gas is most readily detected by using the \ion{Mg}{2}
$\lambda$ 2800\AA\ doublet and the high ionization state gas from the
\ion{C}{4} $\lambda$ 1550\AA\ doublet (Ly$\alpha$ is a special case).  
It is important
that intrinsic features of the quasar not be confused with the
features being searched for, so the Lyman alpha forest should not
be shifted up to the wavelengths of these lines in the rest frame
of the target galaxy or cluster.  For low redshift clusters or
galaxies, this places the restriction that z$_{\rm qso}  < 1.3$ to search
for the \ion{Mg}{2} lines and z$_{\rm qso} < 0.27$ to search for 
\ion{C}{4} absorption.

     This requirement on the redshift favors a quasar population
whose mean redshift is $<$ 1.  Optically selected quasars have a
mean redshift of {$\sim$}1.5 (Hewett et al.\ 1995, Boyle et al.\ 1990) 
while radio selected quasars have a
lower mean redshift of {$\sim$}1.1 (Drinkwater et al.\ 1997), 
but X-ray selected quasars have a mean
redshift of {$\sim$}0.4 (Maccacaro et.\ al.\ 1991) to {$\sim$}0.6 
(Page et al.\ 1996),
depending on the sensitivity of the survey.  
Even in deep X-ray surveys, the mean redshift of the X-ray selected 
quasars is only {$\sim$}1.1 (Bower et al.\ 1996).
Also, relatively few random optical sources are
quasars, but approximately half of all X-ray point sources between energies 
of 0.5 and 2 keV are
quasars (Boyle et al.\ 1994, 1997, Voges et al.\ 1996, Zhao et al.\ 1997).  
Furthermore, because there is a correlation of X-ray
flux with optical and ultraviolet flux (Puchnarewicz et al.\ 1996, Walter and 
Fink 1993, Laor et al.\ 1994), X-ray selected quasars
are preferentially bright in the ultraviolet region where the
absorption line measurements will be made.

     Our strategy has been to use X-ray images, largely from the
ROSAT archives, of galaxies and clusters that are of interest.  We search
these fields for X-ray point sources and eliminate any sources
that correspond to foreground Galactic stars, and in the clusters
of galaxies, to a known cluster member.  Subsequently, we try to
identify the optical counterparts to these X-ray sources on the
Digitized Sky Survey.  There is a practical lower limit to the
brightness of a useful quasar in that it must not require an
inordinate amount of HST to detect absorption lines; this leads
to $V < 19.5$ mag. X-ray targets whose optical counterpart is
fainter than this limit are eliminated, and this removes about
half of the potential targets.  Typically there are $\sim$1--2 good
X-ray targets per ROSAT field searched.

     The error circle associated with a particular X-ray
detection depends on the instrument used, the characteristics of
the particular field, and the strength of the X-ray point source. 
Near the field center, the resolution of the two ROSAT
instruments are 25\arcsec\ FWHM (PSPC) and 5\arcsec\ FWHM (HRI), and since it
is possible to find the centroid to some fraction of the
instrumental resolution, the centers can be located to about 10\arcsec\
for the PSPC and 2\arcsec\ for the HRI.  However, the absolute positions
given by the aspect solution for a particular field can be
incorrect by as much as 15\arcsec, although more typical values are 
5-10\arcsec.  
When there is an identified source in the field, such as a
star, the position of the other sources can be corrected, leading
to error circles of radius 10\arcsec\ for the PSPC and 5\arcsec\ for the HRI. 
When no source can be identified to correct the astrometry, we
use an error circle of radius 20\arcsec, which is conservative.  With
error circles of 20\arcsec, there is usually no more than one object
brighter than 19.5 mag within the circle.  Once a target has been
identified in this manner, the success rate of finding new
quasars spectroscopically is about 50\%, but their redshift may be
too high or uncertain because only one emission line is seen, 
or their magnitude may have decreased below the limit. 
Approximately 1/4 sources that are observed with optical
spectroscopy become acceptable targets for absorption line
studies (1/8 of initial X-ray detections). 

     The angular distance on the sky between the galaxy or
cluster being studied and the quasar depends upon the surface
brightness of the emission from the cluster as well as on the
exposure time of the X-ray field.  For typical PSPC or HRI
exposure times of 8-20 ksec, there are 2-3 quasars with z $<$ 1.3 
in 1000 square arcminutes (Page et al.\ 1996), most brighter than 19.5 mag.
This is somewhat higher than optical searches, which typically find $\sim$1 
quasar in 1000 square arcminutes brighter than 19.5 magnitudes, assuming 
\bv $= 0.3$ (Boyle et al.\ 1990).  
However, for clusters of galaxies, the surface brightness of the
cluster X-ray emission is fairly high and it is difficult to
identify weak X-ray sources (potential background quasars)
against this emission.  Only the brighter X-ray quasars are
visible through the inner regions of a cluster of galaxies, and
for these identifications, the superior angular resolution of the
HRI make these images more useful than the PSPC images, even
though the HRI sensitivity is lower.  Further from the cluster
center ($> 0.5$ Mpc), where the cluster X-ray surface brightness is
substantially lower, it is possible to identify fainter X-ray
point sources for optical follow-up studies.  In practice, for a
typical bright X-ray cluster at a redshift of 0.05, the X-ray
surface brightness is high in the inner 5-10\arcmin, although we have
been able to find quasars as close as 3\arcmin\ from the X-ray center of
a cluster.  

Similar considerations apply to elliptical galaxies, except
that the region of high surface brightness is smaller (1-2\arcmin\ ) and
many ellipticals are not particularly bright X-ray sources.  In
these cases, it has been possible to find background quasars
within 2\arcmin\ of the center of the galaxy, although that was a 
exceptional case.
For our sample, the median distance between 
an object, either the X-ray center of a cluster or an elliptical galaxy, 
and the closest detected quasar is 6-7\arcmin.

\section{The Identification of New Quasars}

\subsection{The Sample}

Our ongoing program searches for X-ray sources behind ellipticals,
groups of galaxies, and clusters of galaxies.   
Our technique
for the identification of background quasars has evolved and
improved since we began this program and we do not claim that our
search was complete to a particular X-ray flux or signal-to-noise
limit.  However, we have searched for quasar identifications for
all 3$\sigma$ X-ray sources within 8\arcmin\ of the cluster centers and all
strong X-ray sources within 12\arcmin\ of the cluster center.  Also, we
investigated all 3$\sigma$ X-ray point sources within the optical
radius (D$_{25}$) of the centers of the elliptical galaxies.

\subsection{Observations}

   Spectroscopic observations were made at Michigan-Dartmouth-MIT Observatory
on the 2.4 m telescope.  The first run was from 4 - 6 March 1994 and used the
Mark III spectrograph with a 2048${\rm x}$2048 thick CCD.  We used a 300 l/mm
grism with a blaze angle of 6400\AA\ and a 1{\farcs}17 slit.  The scale is
0.488 {\arcsec}/pix for this detector, giving a nominal resolution of 
{$\sim$}8\AA\ 
and a spectral range from 4260 - 11070{\AA}.
The second run was from 1 - 3 December 1994 and the third run was from 
17 - 22 May 1995.  Both runs used the Mark III spectrograph with a 
1024${\rm x}$1024 thinned CCD.  We used a 300 l/mm grism with a blaze angle of
5400\AA\ and a 1{\farcs}17 slit.  The scale is
0.781 {\arcsec}/pix for this detector, giving a nominal resolution of 
{$\sim$}8\AA\ 
and a spectral range from 3620 - 9040\AA\ (December 1994) and 3860 - 9352 \AA\
(May 1995). Spectrophotometric standards were selected from 
Massey et.\ al.\ (1988) and Massey and Gronwell (1990).  
Images were bias subtracted
and then flat-fielded using quartz lamp flats.  An illumination correction was
applied using dusk sky flats.  Wavelength solutions were determined from Hg-Ne
lamps, and no significant shift as a function of position on the sky was seen.

     The X-ray sources identified as quasars are listed in Table
1. Column 1 lists the name, column 2 lists the 
foreground or background galaxy or cluster, 
columns 3 and 4 list the right ascension and declination in J2000, column 5 
lists the quasar redshift, column 6 lists the spectroscopic 
visual magnitude, column 7 lists the
date of the optical observation, column 8 lists the 
distance from the galaxy or cluster (X-ray center) in arcminutes, and column 9
lists the X-ray count rates for the detected quasars 
in $10^{-3}$ counts s$^{-1}$ (vignetting corrected) when available.  
The X-ray count rates were 
determined from the High Energy Astrophysics Science Archive Research Center 
(HEASARC) searches of the ROSAT Catalog PSPC RX MPE Sources (HEASARC\_ROSATSRC)
and ROSAT Catalog PSPC WGA Sources (HEASARC\_WGACAT), or were calculated from 
archived data by one of us (J.\ Bregman).  In the cases where the
source was detected in both catalogs (QSO0152$-$137, TON1480), the reported 
count rate is an average of the two given count rates.  
Sources without available X-ray counts 
were either taken from the Hewitt-Burbidge QSO catalog (1993, QSO0020$+$287 and
QSO1124$+$271), discovered serendipitously in the spectrum of another source
(QSO0109$-$153), or located more than 20 arcseconds from the X-ray source and
therefore probably not associated with the X-ray source (QSO1201$+$580).  
For
completeness, all quasars are listed, not just those in the
proper magnitude or redshift range.  It should be noted that the visual
magnitudes listed are of varying accuracy, with those noted as definitely
nonphotometric to be considered as upper limits.  Furthermore, since the 
standards were taken through a narrow slit, the magnitudes are relative, 
not absolute.  For a small number of
sources which have also been imaged in $V$, the mean difference in $V$
magnitude between the spectroscopic magnitude and photometric magnitude
is 0.39 magnitudes.  In the one case of an object whose spectrum was
taken on a photometric night, the difference was 0.23 magnitudes.  
%In all cases, the $V$ magnitudes from the spectra {\it underestimate} the
%true $V$ flux.  
As a cautionary note, however, all the photometric
data were taken at least a year after the spectroscopic data, and QSOs
are known to vary by 0.1-0.4 magnitudes over weeks to years (Trevese et al.\ 
1989, Kaspi 1997, Dietrich et al.\ 1997, Qian and Tao 1997), 
so the true errors are uncertain.

\begin{table}
\dummytable\label{qso_table}
\end{table}

Finding charts are shown for
the quasars in Figure 1.  Each chart is 5\arcmin\ on a side, and north is
up and east is to the left.  The quasar is indicated by an arrow.
The discovery spectra of the quasars are shown in Figure 2.  Major emission
lines are indicated.  NS indicates a poorly subtracted night sky line.

     The sources which were not identified as quasars are listed in Table
2. Column 1 lists the name.  (In this case we have identified each object by
``OBJxx{$\pm$}yy'', where xx and yy are the full right ascension and 
declination.  This was done because many of the observations are of objects
only a few arcseconds apart.)  Column 2 lists the
corresponding foreground or background galaxy or cluster,
columns 3 and 4 list the right ascension and declination in J2000, column 5
lists the date of the observation, and column 6
lists the source identification, if any.  A search of the NASA Extragalactic
Database\footnotemark\ indicated that none of the non-AGN galaxies found in the course of
this work were previously cataloged.  Thus, we have also identified 8 new
galaxies, most of which are probably previously unknown cluster members, based 
on their measured redshifts.  

\begin{table}
\dummytable\label{notqso_table}
\end{table}

\footnotetext{The NASA/IPAC Extragalactic Database (NED)
is operated by the Jet Propulsion Laboratory, California Institute
of Technology, under contract with the National Aeronautics and Space
Administration.}

\section{Conclusions}

We have identified a successful process for locating quasar
candidates behind individual galaxies and clusters of galaxies.
Using X-ray images from archival data, we search for QSO candidates
which are then identified via optical spectroscopy.  We have presented
16 new quasars and galaxies with active galactic nuclei (AGNs), 
and confirmed 5 known quasars and AGNs.  We also
presented 3 QSO/AGN candidates with uncertain redshift identifications.
We find that approximately 1/8 of X-ray identified sources prove to be
QSOs with the correct redshift, spectral shape, and magnitude to be
used in HST studies of the instellar and intergalactic media of galaxies
and clusters of galaxies.

\acknowledgements

This research has made use of data obtained from the High Energy Astrophysics Science Archive
Research Center (HEASARC), provided by NASA's Goddard Space Flight Center, 
and of the Digitized Sky Survey, as
provided by the Space Telescope Science Institute.  
%In addition,
%this research has made use of the NASA/IPAC Extragalactic Database (NED)   
%which is operated by the Jet Propulsion Laboratory, California Institute   
%of Technology, under contract with the National Aeronautics and Space      
%Administration.                                                            
We are indebted to 
the staff at the Michigan-Dartmouth-MIT Observatory for their
assistance and support in this matter.  Also, we would like to
thank Ken Banas for his assistance in identifying X-ray point
sources in some of the fields.  
Finally, we wish to acknowledge support from NASA
grants NAGW-2135 and NAGW-4448.

\newpage
%\begin{figure}
\figurenum{1a}
%\plotone{Knezek.fig1a_2.eps}
%\caption{Finding charts for detected emission line objects.  North is up and east is to the left.  All charts are 5 arcminutes on a side.  The object is indicated by an arrow.}
%\end{figure}
\figcaption[knezek.fig1a.ps]{Finding charts for detected emission line objects.  North is up and east is to the left.  All charts are 5 arcminutes on a side.  The object is indicated by an arrow. \label{fig1a} }

%\newpage
%\begin{figure}
\figurenum{1b}
%\plotone{Knezek.fig1b_2.eps}
%\caption{Finding charts for detected emission line objects.  North is up and east is to the left.  All charts are 5 arcminutes on a side.  The object is indicated by an arrow.}
%\end{figure}
\figcaption[knezek.fig1b.ps]{Finding charts for detected emission line objects.  North is up and east is to the left.  All charts are 5 arcminutes on a side.  The object is indicated by an arrow. \label{fig1b} }

%\newpage
%\begin{figure}
\figurenum{1c}
%\plotone{Knezek.fig1c_2.eps}
%\caption{Finding charts for detected emission line objects.  North is up and east is to the left.  All charts are 5 arcminutes on a side.  The object is indicated by an arrow.}
%\end{figure}
\figcaption[knezek.fig1c.ps]{Finding charts for detected emission line objects.  North is up and east is to the left.  All charts are 5 arcminutes on a side.  The object is indicated by an arrow. \label{fig1c} }

%\begin{figure}
\figurenum{1d}
%\plotone{Knezek.fig1d_2.eps}
%\caption{Finding charts for detected emission line objects.  North is up and east is to the left.  All charts are 5 arcminutes on a side.  The object is indicated by an arrow.}
%\end{figure}
\figcaption[knezek.fig1d.ps]{Finding charts for detected emission line objects.  North is up and east is to the left.  All charts are 5 arcminutes on a side.  The object is indicated by an arrow. \label{fig1d} }

%\newpage
%\begin{figure}
\figurenum{2a}
%\plotone{Knezek.fig2a.eps}
%\caption{ Discovery spectra for detected emission line objects.  Major emission lines are indicated in each spectrum.  NS indicates a poorly subtracted night sky line.}
%\end{figure}
\figcaption[knezek.fig2a.ps]{ Discovery spectra for detected emission line objects.  Major emission lines are indicated in each spectrum.  NS indicates a poorly subtracted night sky line. \label{fig2a}}

%\newpage
%\begin{figure}
\figurenum{2b}
%\plotone{Knezek.fig2b.eps}
%\caption{ Discovery spectra for detected emission line objects.  Major emission lines are indicated in each spectrum.  NS indicates a poorly subtracted night sky line.}
%\end{figure}
\figcaption[knezek.fig2b.ps]{ Discovery spectra for detected emission line objects.  Major emission lines are indicated in each spectrum.  NS indicates a poorly subtracted night sky line. \label{fig2b}}
 
%\newpage
%\begin{figure}
\figurenum{2c}
%\plotone{Knezek.fig2c.eps}
%\caption{ Discovery spectra for detected emission line objects.  Major emission lines are indicated in each spectrum.  NS indicates a poorly subtracted night sky line.}
%\end{figure}
\figcaption[knezek.fig2c.ps]{ Discovery spectra for detected emission line objects.  Major emission lines are indicated in each spectrum.  NS indicates a poorly subtracted night sky line. \label{fig2c}}

\end{document}